\documentclass[letter]{aa}

\usepackage{graphicx}
\usepackage{txfonts}
\begin{document} 

\title{The Kormendy relation of early-type galaxies as a function of wavelength in Abell S1063, MACS J0416.1-2403, and MACS J1149.5+2223\thanks{The table of structural parameters described in Appendix A is only available at the CDS via anonymous ftp to cdsarc.cds.unistra.fr (130.79.128.5) or via https://cdsarc.cds.unistra.fr/viz-bin/cat/J/A+A/671/L9}}


\author{L. Tortorelli \inst{1} \and A. Mercurio \inst{2,3} \and G. Granata \inst{4,5} \and P. Rosati \inst{6} \and C. Grillo \inst{4,5} \and M.  Nonino \inst{7} \and A.  Acebron \inst{4,5} \and G. Angora \inst{3,6} \and P.  Bergamini \inst{4,8} \and G. B.  Caminha \inst{9,10} \and U. Me\v{s}tri\'{c} \inst{8} \and E.  Vanzella \inst{8} }

\institute{University Observatory, Faculty of Physics, Ludwig-Maximilians-Universit\"at M\"unchen, Scheinerstr. 1, 81679 Munich, Germany\\ \email{Luca.Tortorelli@physik.lmu.de} \and
Dipartimento di Fisica “E.R. Caianiello”,  Universit\`a Degli Studi di Salerno, Via Giovanni Paolo II, I–84084 Fisciano (SA), Italy\\ 
\email{amata.mercurio@inaf.it} \and
 INAF-Osservatorio Astronomico di Capodimonte, Salita Moiariello 16, I-80131 Napoli, Italy \and 
Dipartimento di Fisica, Universit\`a degli Studi di Milano, Via Celoria 16, I-20133 Milano, Italy \and
 INAF - IASF Milano, via Corti 12, I-20133 Milano, Italy \and
Dipartimento di Fisica e Scienze della Terra,  Universit\`a di Ferrara,  Via Saragat 1, 44122 Ferrara, Italy \and
INAF - Osservatorio Astronomico di Trieste,  via Tiepolo 11, 34143, Trieste, Italy \and
INAF - OAS, Osservatorio di Astrofisica e Scienza dello Spazio di Bologna, via Gobetti 93/3, I-40129 Bologna, Italy \and
Technical University of Munich, TUM School of Natural Sciences, Department of Physics, James-Franck-Str 1, 85748 Garching, Germany \label{tum} \and
Max-Planck-Institut f\"ur Astrophysik, Karl-Schwarzschild-Str. 1, D-85748 Garching, Germany \label{mpa}}

   \date{Received XXXX; accepted XXXX}


  \abstract
   {The wavelength dependence of the projection of the fundamental plane along the velocity dispersion axis, namely the Kormendy relation,  is well characterised at low redshift but poorly studied at intermediate redshifts.  The Kormendy relation provides information on the evolution of the population of early-type galaxies (ETGs).  Therefore,  by studying it,  we may shed light on the assembly processes of these objects and their size evolution.  As studies at different redshifts are generally conducted in different rest-frame wavebands,  it is important to investigate whether the Kormendy relation is dependent on wavelength. Knowledge of such a dependence   is fundamental to correctly interpreting the conclusions we might draw from these studies.}
   {We analyse the Kormendy relations of the three Hubble Frontier Fields clusters, Abell\,S1063 at z = 0.348, MACS\,J0416.1-2403 at z = 0.396, and MACS\,J1149.5+2223 at z = 0.542, as a function of wavelength. This is the first time the Kormendy relation of ETGs has been explored consistently over such a large range of wavelengths at intermediate redshifts.}
   {We exploit very deep Hubble Space Telescope photometry, ranging from the observed \textit{B}-band to the \textit{H}-band,  and VLT/MUSE integral field spectroscopy. We improve the structural parameter estimation we performed in a previous work by means of a newly developed \textsc{python} package called \textsc{morphofit}.}
   {With its use on cluster ETGs,  we find that the Kormendy relation slopes increase smoothly with wavelength from the optical to the near-infrared (NIR) bands in all three clusters,  with the intercepts becoming fainter at lower redshifts due to the passive ageing of the ETG stellar populations.  The slope trend is consistent with previous findings at lower redshifts.}
   {The slope increase with wavelength implies that smaller ETGs are more centrally concentrated than larger ETGs in the NIR with respect to the optical regime. As different bands probe different stellar populations in galaxies, the slope increase also implies that smaller ETGs have stronger internal gradients with respect to larger ETGs. }


\titlerunning{The Kormendy relation as a function of wavelength}

\maketitle
%

\section{Introduction}

Studies of how the structural parameters of galaxies are linked to their dynamics and stellar properties via scaling relations and how these change as a function of redshift, environment, and wavelength have been used many times as probes to understand the formation and evolution of galaxies.  The fundamental plane (FP; \citealt{Dressler1987,Djorgovski1987,Bender1992,Saglia1993,Donofrio2022,Donofrio2023}) and its projections are among the most important galaxy scaling relations.  The FP links the structural properties of early-type galaxies (ETGs), such as their surface brightnesses and effective radii, with their dynamics.  One of its projections is the Kormendy relation (KR; \citealt{Kormendy1977}). The KR links the effective radius $R_\mathrm{e}$ of an ETG to its average surface brightness within that radius $\left \langle \mu \right \rangle_{\mathrm{e}}$, $\left \langle \mu \right \rangle_{\mathrm{e}} = \alpha + \beta \log{R_{\mathrm{e}}}$. The KR is a physical correlation,  presumably reflecting the difference in the origin of bright and faint ellipticals and bulges; it provides information on how the size and light distribution in ETGs evolve as a function of redshift. This implies that the KR can be used to understand whether ETGs have completed their growth in mass and size at their redshift or if there is still significant growth up to $z=0$.  Furthermore, the KR can also be used to study the bulge formation of disk galaxies \citep{Gadotti2009}: classical bulges can be disentangled from pseudobulges, the latter being $3\sigma$ outliers with respect to the best-fitting KR \citep{Costantin2018,Gao2020}.

The KR has been studied extensively in the literature \citep{Ziegler1999,LaBarbera2003,Longhetti2007,Rettura2010,Saracco2014}. However, the conclusions drawn from it are contrasting; in particular, they depend on how the sample of galaxies is selected, that is, whether this selection is based on colours, S\'ersic indices, or stellar populations (e.g. \citealt{Saracco2010,Andreon2016,Fagioli2016}). In \citealt{Tortorelli2018},  we investigated the effect of sample selection and found that a selection purely based on colours may bias the KR parameter estimates,  while galaxies selected through S\'ersic indices (ETGs), visual inspection (ellipticals),  and spectra (passives) constitute a more coeval population.  

Additionally, the morphologies and therefore the sizes of galaxies may appear different at different wavelengths. As different wavelengths probe different regions and physical processes inside the galaxies, the KR does not necessarily hold or have the same parameters at all wavelengths. The surface brightness in the near-infrared (NIR) is less affected by gas and dust extinction; it is instead dominated by the older stellar population of galaxies, which constitutes their main stellar mass component, especially that of ETGs. Therefore, the contrast between the NIR and the optical observations translates into the contrast between the underlying mass component (older stellar population) and the younger stellar population component. Therefore, the wavelength range used to study the KR may impact the conclusions that we can draw from it. Studies of the wavelength dependence of the KR  at low redshift  have been conducted by \citealt{LaBarbera2010}, for example. These studies show that there is evolution of the KR slope as a function of wavelength for Sloan Digital Sky Survey (SDSS) ETGs selected via colours in the form of a steepening of the relation from the $\textit{g}$ to the $\textit{K}$ observed bands. However, studies at higher redshifts ($z > 0.3$) with samples of ETGs consistently selected at different redshifts are still missing.

In this letter, we use data from the Hubble Space Telescope (HST) Frontier Fields (FF) survey \citep{Lotz2017} to study the KR of ETGs as a function of wavelength in three clusters at intermediate redshift, namely Abell\,S1063 at $z = 0.348$, MACS\,J0416.1-2403 at $z = 0.396$, and MACS\,J1149.5+2223 at $z = 0.542$. We build the KR for ETGs selected via S\'ersic indices only, following the conclusions in \citealt{Tortorelli2018} regarding the consistent selection of ETGs at different redshifts.  

We measured the structural parameters using a new \textsc{python} package we developed called \textsc{morphofit} \citep{Tortorelli2023}. We used the data-analysis pipeline first introduced in \citealt{Tortorelli2018} and refined in \citealt{Tortorelli2023}, which involves fitting the surface-brightness profiles of galaxies in images of increasing size. 

The letter is structured as follows. In Sect. \ref{section:dataset}, we describe the photometric and spectroscopic data used for this analysis. In Sect. \ref{section:struct_param_estimate}, we summarise the structural parameter estimation process using \textsc{morphofit}. In Sect. \ref{section:kormendy_wave}, we present our results for the Kormendy-relation behaviour as a function of wavelength. Finally, we provide our main conclusions in Sect. \ref{section:conclusions}. 

Unless otherwise stated, we give errors at the $68$ per cent confidence level,  and we report the circularised effective radii. Throughout this paper, we use $H_0 = 70 \mathrm{km\ s^{-1}\ Mpc^{-1}}$ in a flat cosmology with $\Omega_{\mathrm{M}}$ = 0.3 and $\Omega_{\Lambda}$ = 0.7. In the adopted cosmology, $1\arcsec$ corresponds to $4.921\ \mathrm{kpc}$ at $z = 0.348$, to $5.340\ \mathrm{kpc}$ at $z = 0.396,$ and to $6.364\ \mathrm{kpc}$ at $z = 0.542$. 

\section{Dataset}
\label{section:dataset}

We analysed the three clusters Abell S1063 (AS1063) at $z = 0.348$, MACS J0416.1-2403 (M0416) at $z = 0.396$, and MACS J1149.5+2223 (M1149) at $z = 0.542$. A wealth of multi-wavelength and wide-field data are available for AS1063, M0416, and M1149  from photometry and spectroscopy \citep{Rosati2014,Karman2015,Grillo2016,Mercurio2021}. The photometric data we use in this study are available at the STScI Mikulski Archive for Space Telescopes (MAST) \footnote{https://archive.stsci.edu/prepds/frontier/}. These data belong to the FF programme\footnote{http://www.stsci.edu/hst/campaigns/frontier-fields/HST-Survey} (PI: J. Lotz, \citealt{Lotz2017}), which was designed to combine the power of \textit{HST} and \textit{Spitzer} with the natural gravitational telescope effect of massive high-magnification clusters of galaxies. These datasets allow us, for instance, to test predictions of the $\Lambda \mathrm{CDM}$ model \citep{Annunziatella2017,Sartoris2020}, to measure the Hubble constant value \citep{Grillo2018,Grillo2020}, to refine weak and strong lensing models in order to map the total mass distribution in clusters \citep{Gruen2013,Caminha2016,Bergamini2019,Granata2022}, and to serendipitously discover very distant lensed galaxies up to $z \sim 6$ \citep{Vanzella2016,Balestra2018}.

In order to measure the structural parameters (i.e. effective radii and surface brightnesses), we used the $0.060\ \mathrm{arcsec/pixel}$ images in all seven optical/NIR bands of the FF programme. The three optical bands \textit{F435W, F606W, F814W} belong to the Advanced Camera for Surveys (ACS), while the four NIR bands \textit{F105W, F125W, F140W, F160W} belong to the Wide Field Camera 3 (WFC3) IR imager. These bands span a wavelength range from $\sim 3500$ \AA , corresponding to the observed Johnson \textit{B}-band filter, to $\sim 17400$ \AA, corresponding to the observed \textit{H} filter. This roughly corresponds to the rest-frame ranges: \textit{u} to \textit{J} band for AS1063 and M0416, and near-UV to \textit{Y} band for M1149. We use the \textit{drz} science images, the \textit{rms} images, and the \textit{exp} exposure-time-map images.

\begin{table*}[ht!]
\small
\centering
\begin{tabular}{l c c c c c c c c c c}
\hline
\hline
&&\textbf{AS1063}&&&\textbf{M0416}&&&\textbf{M1149}&\\
\hline
& $\alpha$ & $\beta$ & $\sigma$ & $\alpha$ & $\beta$ & $\sigma$ & $\alpha$ & $\beta$ & $\sigma$\\
\textit{F435W} & 20.97 $\pm$ 0.11 & 3.11 $\pm$ 1.08 & 0.65 $\pm$ 0.01 & 21.04 $\pm$ 0.13 & 2.81 $\pm$ 0.69 & 0.70 $\pm$ 0.01 & 20.21 $\pm$ 0.66 & 4.69 $\pm$ 2.14 & 0.94 $\pm$ 0.01\\
\textit{F606W} & 19.12 $\pm$ 0.11 & 3.86 $\pm$ 0.61 & 0.72 $\pm$ 0.01 & 19.06 $\pm$ 0.11 & 2.85 $\pm$ 0.34 & 0.65 $\pm$ 0.01 & 19.00 $\pm$ 0.29 & 2.70 $\pm$ 1.02 & 0.50 $\pm$ 0.01\\
\textit{F814W} & 18.19 $\pm$ 0.11 & 3.72 $\pm$ 0.54 & 0.74 $\pm$ 0.01 & 18.04 $\pm$ 0.11 & 3.29 $\pm$ 0.38 & 0.74 $\pm$ 0.01 & 17.56 $\pm$ 0.22 & 3.27 $\pm$ 0.83 & 0.63 $\pm$ 0.01\\
\textit{F105W} & 17.72 $\pm$ 0.12 & 4.30 $\pm$ 0.64 & 0.81 $\pm$ 0.01 & 17.56 $\pm$ 0.10 & 3.38 $\pm$ 0.48 & 0.81 $\pm$ 0.01 & 16.85 $\pm$ 0.35 & 3.74 $\pm$ 1.80 & 0.72 $\pm$ 0.01\\
\textit{F125W} & 17.47 $\pm$ 0.13 & 4.31 $\pm$ 0.72 & 0.82 $\pm$ 0.01 & 17.18 $\pm$ 0.12 & 3.60 $\pm$ 0.67 & 0.85 $\pm$ 0.01 & 16.39 $\pm$ 0.72 & 4.81 $\pm$ 3.46 & 0.84 $\pm$ 0.01\\
\textit{F140W} & 17.32 $\pm$ 0.13 & 4.22 $\pm$ 0.75 & 0.83 $\pm$ 0.01 & 17.07 $\pm$ 0.13 & 3.76 $\pm$ 0.68 & 0.89 $\pm$ 0.01 & 16.06 $\pm$ 0.27 & 5.45 $\pm$ 1.39 & 0.86 $\pm$ 0.01\\
\textit{F160W} & 17.21 $\pm$ 0.14 & 5.28 $\pm$ 0.87 & 0.91 $\pm$ 0.01 & 16.85 $\pm$ 0.11 & 3.62 $\pm$ 0.54 & 0.81 $\pm$ 0.01 & 16.00 $\pm$ 0.29 & 5.08 $\pm$ 1.58 & 0.86 $\pm$ 0.01\\
\hline
\hline
\end{tabular}
\caption{The table reports the best-fitting slopes,  intercepts,  observed scatters, and their $1\sigma$ errors obtained by fitting the KR to the ETG samples. }
\label{table:KR_parameters}
\end{table*}

The AS1063 and M1149 spectroscopically confirmed cluster members we use in our analysis are the same as those selected in \citealt{Tortorelli2018} using the Multi Unit Spectroscopic Explorer (MUSE) spectrograph integral field unit (IFU) spectra.  The confirmed members are $95$ for AS1063, which have redshifts in the range $0.335 \le z \le 0.361$ \citep{Karman2015}, and $68$ for M1149, which have redshifts of $0.52 \le z \le 0.57$ \citep{Grillo2016}. \textit{HST} imaging and \textit{MUSE} IFU spectra allow us to reach a completeness in apparent magnitude of 1.0 down to a value of $22.5$ in the \textit{F814W} waveband \citep{Caminha2016}. This limit roughly corresponds to a stellar mass value of $M_{*} \sim 10^{9.8} \mathrm{M_{\odot}}$ and $M_{*} \sim 10^{10.0} \mathrm{M_{\odot}}$ for AS1063 and M1149, respectively, for the typical spectral energy distribution of the sources we are interested in and considering a Salpeter initial mass function (IMF; \citep{Salpeter1955}).  The cluster members for M0416 are also those for which MUSE spectra are available; there are $119$ members, which have redshifts of $0.38 \le z \le 0.41$. The description of the member selection is detailed in \citealt{Annunziatella2017} and \citealt{Caminha2017}.

In \citealt{Tortorelli2018}, we defined and compared four different galaxy samples according to (a) S\'ersic indices: early-type (`ETG'), (b) visual inspection: `ellipticals', (c) colours: `red', (d) spectral properties: `passive'. We showed that the KR built with the `ETG' sample is fully consistent with the ones obtained with the `elliptical' and `passive' samples. On the other hand, the KR slope built with the `red' sample is only marginally consistent with those obtained with the other samples. Therefore, in this work, we analyse the results using only the sample of ETGs selected via S\'ersic indices.

\section{Structural parameter estimates with \textsc{morphofit}}
\label{section:struct_param_estimate}

In order to build the KR, we need to estimate the structural parameters of spectroscopically confirmed cluster members. We do this using the \textsc{python} package \textsc{morphofit}\footnote{https://pypi.org/project/morphofit/} \citep{Tortorelli2023}. The package uses \textsc{SExtractor} \citep{Bertin1996} and \textsc{GALFIT} \citep{Peng2011} to automatically fit the surface-brightness profile of a set of user-defined galaxies. The code is highly parallelisable, making it suitable for modern wide-field photometric surveys. A complete description of the software features can be found in \citealt{Tortorelli2023}.

The methodology for the estimation of structural parameters is a refined version of that highlighted in \citealt{Tortorelli2018}. To deal with the cluster-crowded environment and the intracluster light contribution, we adopt an iterative approach that analyses images of increasing size (from stamps to the full images) using different point-spread-function (PSF) images, background-estimation methods, and sigma image creation.

The analysis with \textsc{morphofit} starts by running \textsc{SExtractor} in forced photometry mode on the drizzled images of the three clusters in all seven wavebands. The structural parameters estimated with \textsc{SExtractor} constitute the initial values of the surface-brightness profile fits with \textsc{GALFIT}. As \textsc{GALFIT} requires a PSF image for the light-profile convolution,  and it has been proven that different PSFs may lead to different structural parameter estimates \citep{Vanzella2019},  we built four different PSF images with four different methods for each cluster and each waveband (see Fig. 2 in \citealt{Tortorelli2023}) to average out systematic effects arising from a specific model PSF.  We used the \textsc{SExtractor} catalogue to select stars on the images based on their loci on the magnitude--size (MAG\_AUTO vs FLUX\_RADIUS \textsc{SExtractor} parameters) and magnitude--maximum surface brightness (MAG\_AUTO vs MU\_MAX \textsc{SExtractor} parameters) planes and then cut stamps of $50$ pixels in size around them. The sample of stars is further refined based on a signal-to-noise ratio ($100 \le S/N \le 800$) and isolability criterion (no detected sources around). 

\begin{figure*}[ht!]
\centering
\includegraphics[width=9cm]{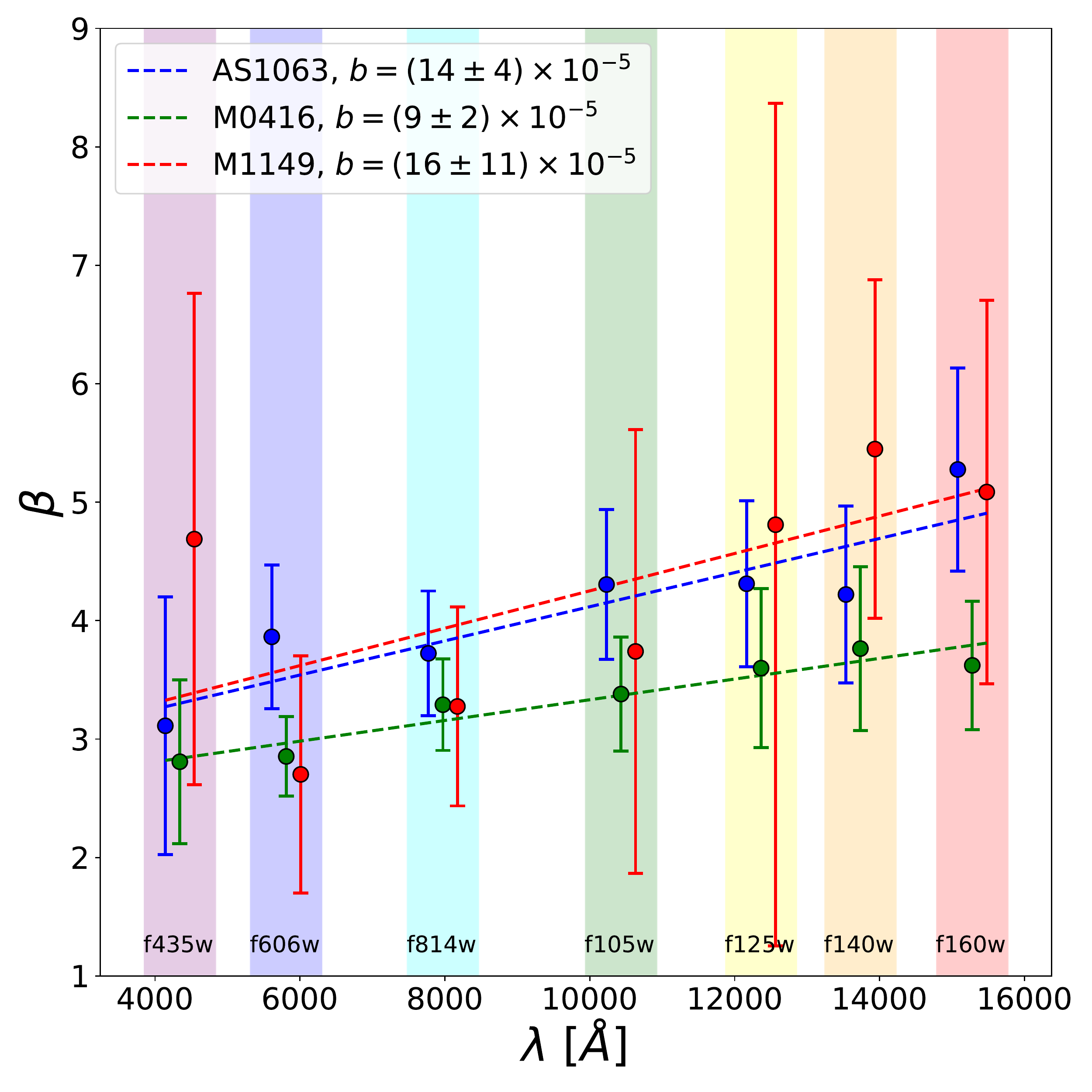}
\includegraphics[width=9cm]{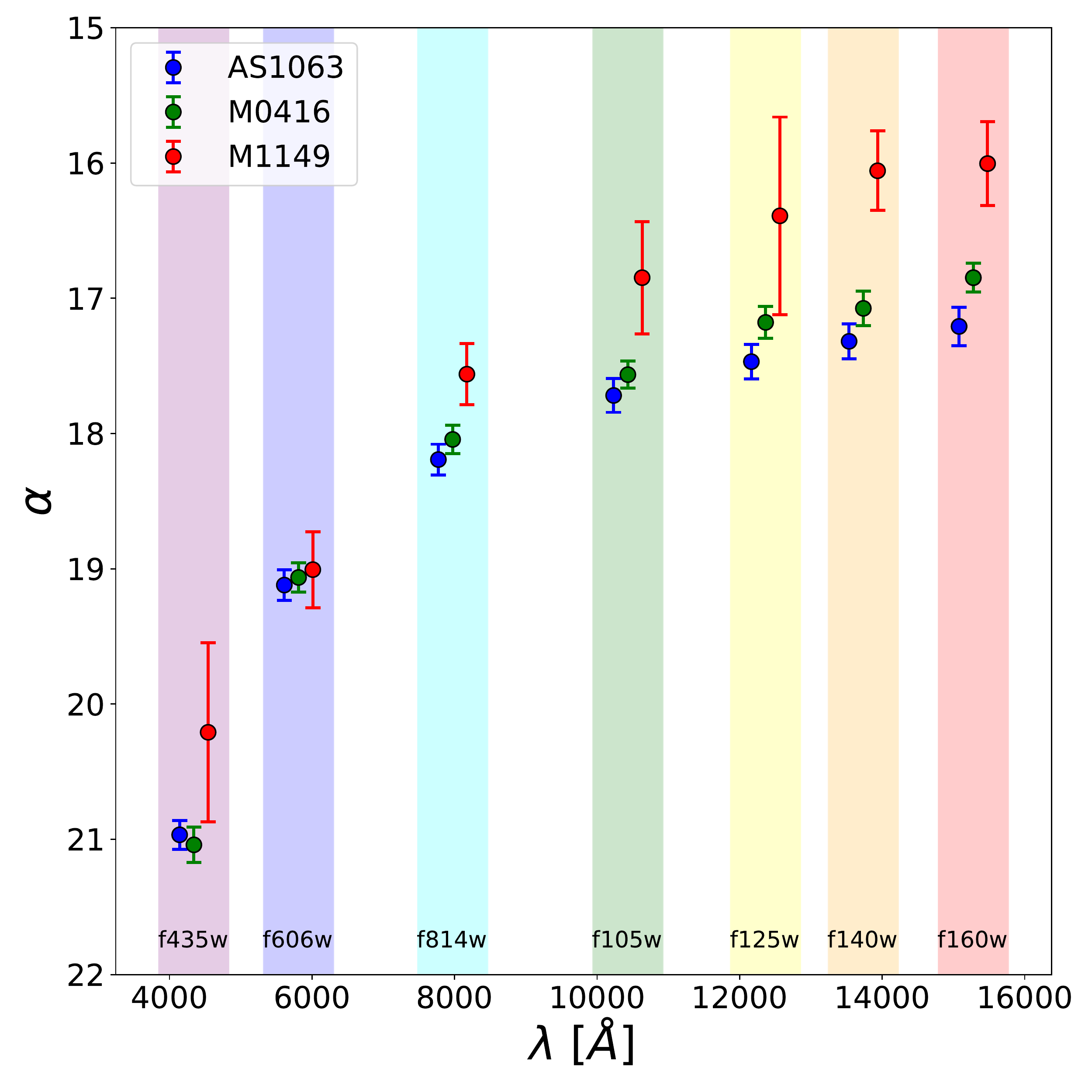}
\caption{Best-fitting KR slope $\beta$ (left panel) and intercept $\alpha$ (right panel) evolution as a function of wavelength for AS1063, M0416, and M1149 (blue, green, and red points, respectively).  The scatter points refer to the best-fitting KR slopes and intercepts obtained from the linear regression on the galaxy structural parameters measured with \textsc{morphofit}. The dashed lines in the left panel represent the best-fitting linear relations that describe the trend of increasing slope values as a function of wavelength for the three clusters. The coloured bands group the scatter points belonging to the same waveband,  which are displaced for clarity. The left plot legend shows the slope values and their $1\sigma$ uncertainties.}
\label{KR_slopes_intercepts_wavelength}
\end{figure*}

\textsc{morphofit} then cuts stamps around the spectroscopically confirmed cluster members and fits their light profiles and those of the neighbouring objects in the stamps with \textsc{GALFIT}. We fit single S\'ersic profiles with initial values of the positions, magnitudes, half-light radii, axis ratios, and position angles given by the \textsc{SExtractor} parameters XWIN\_IMAGE, YWIN\_IMAGE, MAG\_AUTO, FLUX\_RADIUS, BWIN\_IMAGE/AWIN\_IMAGE, and THETAWIN\_IMAGE, respectively. The initial values of the S\'ersic indices are kept fixed at $2.5$. The fit is performed on each member for all seven wavebands with different combinations of four kinds of PSF images, two background estimate methods and two different ways of providing \textsc{GALFIT} with sigma images (see Sect. 2 and 4 in \citealt{Tortorelli2023} for more details).  For each galaxy,  we combine the measurements $X_i$ obtained with the various combinations (`comb') into a single estimate $X_{\mathrm{est}}$ of the structural parameters via an error-weighted mean:
\begin{eqnarray}
\label{weighted_average_formula}
X_{\mathrm{est}} &=& \frac{ \sum_{i=1}^{N_{\mathrm{comb}}} X_i \ w_i}{\sum_{i=1}^{N_{\mathrm{comb}}} w_i} \ ,  \\
w_i &=& \frac{1}{X_{i,\mathrm{err}}^2} \ ,
\end{eqnarray}
where $w_i$ and $X_{i,\mathrm{err}}$ are the weights and the errors quoted by \textsc{galfit} for each fit,  respectively.   The error on the estimates $\sigma_{\mathrm{est}}$ is computed as the square root of the unbiased, weighted estimator sample variance:
\begin{equation}
\sigma_{\mathrm{est}}^2 = \frac{\sum_{i=1}^{N_{\mathrm{comb}}} w_i}{\left( \sum_{i=1}^{N_{\mathrm{comb}}} w_i \right)^2 - \sum_{i=1}^{N_{\mathrm{comb}}} w_i^2} \sum_{i=1}^{N_{\mathrm{comb}}} w_i \left( X_i - X_{\mathrm{est}} \right)^2 \ .
\end{equation}

These structural parameters are, in turn, used as initial values for the light-profile fit of galaxies in images of  increasing size, first on image regions and then on the full images. We used the same combination of PSF images, background, and sigma images as the fit on stamps. The final structural parameters we used to build the KR at different wavelengths are a weighted mean (equation \ref{weighted_average_formula}) of those obtained with the different combinations in the seven wavebands on the full images. The structural parameters for the three clusters are provided as supplementary material to this letter (Appendix \ref{appendix:cat_description}).

\section{The Kormendy relation as a function of wavelength}
\label{section:kormendy_wave}

We used the structural parameters estimated in section \ref{section:struct_param_estimate} to select the sample of ETGs we used to build the KR. Following \citealt{Tortorelli2018}, we fit galaxies with single S\'ersic profiles and select ETGs as those having S\'ersic indices in the \textit{F814W} waveband, $\mathrm{n_{\textit{F814W}}} \ge 2.5$. Additionally, we limit our analysis to galaxies brighter than the completeness limit, $m_{\mathrm{\textit{F814w}}} \le 22.5\ \mathrm{ABmag}$, and we do not consider BCGs in the fit. The number of galaxies satisfying this criterion is 42, 50, and 35 for AS1063, M0416, and M1149, respectively. 

We performed the linear regression analysis of the KR, $\left \langle \mu \right \rangle_{\mathrm{e}} = \alpha + \beta \log{\mathrm{R}_{\mathrm{e}}}$, as in \citealt{Tortorelli2018}, using the bivariate correlated errors and intrinsic scatter estimator (BCES; \citealt{Akritas1996}) method. $\left \langle \mu \right \rangle_{\mathrm{e}}$ and $R_{\mathrm{e}}$ are in units of $\mathrm{mag\ arcsec^{-2}}$ and $\mathrm{kpc}$, respectively, and the surface brightness is corrected for the cosmological dimming effect.  Table \ref{table:KR_parameters} reports the slopes, intercepts, scatters, and their $1\sigma$ uncertainties of the best-fitting KRs built with the ETG samples for the three clusters in the seven wavebands. 

Figure \ref{KR_slopes_intercepts_wavelength} shows the slope and intercept evolution with wavelength for AS1063, M0416, and M1149. We fit a linear relation to the slopes as a function of wavelength. We find the slopes of the three clusters evolve with wavelength, smoothly increasing their values from the observed \textit{B}-band to the observed \textit{H}-band. This result extends the trend of the KR slope increase with wavelength already found at low redshift by \citealt{LaBarbera2010} to intermediate redshift. In our work, the NIR slopes have higher values and are only marginally consistent within the errors with the values quoted by \citealt{LaBarbera2010}, despite our errors being larger due to the much smaller number of objects. 

A steeper KR implies that the difference in the surface brightness between small and large ETGs is larger than that obtained with a shallower KR. This physically implies that smaller ETGs are more centrally concentrated than larger ETGs in the NIR regime with respect to the optical one. As different wavebands probe different stellar populations in the galaxy, the smooth increase in slope from optical to NIR also implies that smaller ETGs have stronger internal stellar population gradients than galaxies with larger effective radii. Additionally, this result points to the fact that the analysis of the KR at different redshifts should be conducted in the same rest-frame bands; otherwise, the wavelength evolution may impact the conclusions from the KR analysis. 

We also split the samples of cluster members in apparent magnitude bins to replicate the conclusions in \citealt{Nigoche-Netro2008} in regards to the KR slope dependence on the width of the magnitude range and the brightness of galaxies within the magnitude range.  We were not able to assess whether this trend is also present in our data,  given the small sample of spectroscopically confirmed members.  The errors estimated via bootstrap are very high, because with small samples, removing an object changes the KR parameters dramatically.  Therefore, any trend existing with magnitude is completely masked by the very large errors. 

The trend of the intercepts shows an expected behaviour as a function of wavelength. The intercepts of the three clusters get brighter at larger wavelengths, because the sample of ETGs is expected to be constituted by a population of galaxies with passive spectral energy distributions. The trend as a function of redshift is also consistent with what we already found in \citealt{Tortorelli2018}, meaning that the intercepts become fainter at lower redshift due to the passive ageing of the ETG stellar populations. The scatter points of AS1063 and M0416 are almost consistent at all wavelengths, given their smaller difference in redshift with respect to M1149.

\section{Conclusions}
\label{section:conclusions}

In this letter, we present our investigation of how the KR parameters change as a function of the wavelength range probed. We performed this analysis using spectroscopically confirmed cluster members of the three FF clusters, Abell S1063 ($z=0.348$), MACS J0416.1-240($z=0.396$), and MACS J1149.5+2223 ($z=0.542$) in seven photometric wavebands from the observed \textit{B}-band to the observed \textit{H}-band. 

We measured the galaxy structural parameters for the KR using the \textsc{python} package \textsc{morphofit} \citep{Tortorelli2023}, following a refined version of the methodology of increasing image size already adopted in \citealt{Tortorelli2018}.  We used the structural parameters to select ETGs as those with S\'ersic indices, $\mathrm{n_{\textit{F814W}}} \ge 2.5$, and magnitude brighter than the completeness limit, $m_{\mathrm{\textit{F814w}}} \le 22.5\ \mathrm{ABmag}$.

We built the KR across the whole range of available wavelengths, and find that the KR intercepts follow an expected trend, becoming fainter at lower redshift due to the passive ageing of the ETG stellar populations. We also find that the slopes of the KRs increase smoothly with wavelength for all three clusters. 

This result extends the conclusions already found by \citealt{LaBarbera2010} at low redshift with SDSS to intermediate redshifts. The slope increase with wavelength implies that smaller ETGs are more centrally concentrated in the NIR than those with larger radii with respect to the optical regime. As different wavelengths probe different stellar populations, this also implies that smaller ETGs have stronger internal stellar population gradients than larger ETGs. Our investigation of the slope change with wavelength suggests that studies addressing the KR evolution should be conducted at similar rest-frame wavebands at different redshifts for a robust comparison.

\begin{acknowledgements}
We acknowledge financial contributions by PRIN-MIUR 2017WSCC32 "Zooming into dark matter and proto-galaxies with massive lensing clusters" (P.I.: P.Rosati), INAF ``main-stream'' 1.05.01.86.20: "Deep and wide view of galaxy clusters (P.I.: M. Nonino)" and INAF ``main-stream'' 1.05.01.86.31 "The deepest view of high-redshift galaxies and globular cluster precursors in the early Universe" (P.I.: E. Vanzella). The CLASH Multi-Cycle Treasury Program is based on observations made with the NASA/ESA Hubble Space Telescope. The Space Telescope Science Institute is operated by the Association of Universities for Research in Astronomy, Inc., under NASA contract NAS 5-26555. Based on observations made with the European Southern Observatory Very Large Telescope (ESO/VLT) at Cerro Paranal, under programme IDs 60.A-9345(A), 095.A-0653(A), 294.A-5032(A) and 186.A-0798(A).
\end{acknowledgements}

\bibliographystyle{aa}
\bibliography{tortorelli_et_al_2023_bibliography}


\appendix

\section{Catalogues of structural parameters}
\label{appendix:cat_description}

The catalogues of structural parameters for the three clusters AS1063, M0416, and M1149 are provided as supplementary material to this letter. Catalogues are provided in the form of \textsc{FITS} tables. The table column names (in typewritten font) and their descriptions are as follows:
\begin{itemize}
\item $\tt{ID}$: ID in serial order.
\item  $\tt{RA}$: right ascension in degrees.
\item  $\tt{DEC}$: declination in degrees.
\item  $\tt{MAG\_x}$: total magnitude in the x-band.
\item  $\tt{MAG\_ERR\_x}$: total magnitude error in the x-band.
\item  $\tt{RE\_x}$: circularised effective radius in kpc in the x-band.
\item  $\tt{RE\_ERR\_x}$: circularised effective radius error in kpc in the x-band.
\item  $\tt{MU\_x}$: average surface brightness within the effective radius in $\mathrm{mag\ arcsec^{-2}}$ in the x-band.
\item  $\tt{MU\_ERR\_x}$: average surface brightness within the effective radius error in $\mathrm{mag\ arcsec^{-2}}$ in the x-band.
\item  $\tt{N\_x}$: S\'ersic index in the x-band.
\item  $\tt{N\_ERR\_x}$: S\'ersic index error in the x-band.
\item  $\tt{AR\_x}$: axis ratio in the x-band.
\item  $\tt{AR\_ERR\_x}$: axis ratio error in the x-band.
\item  $\tt{PA\_x}$: position angle in degrees in the x-band.
\item  $\tt{PA\_ERR\_x}$: position angle error in degrees in the x-band.
\end{itemize}

$\tt{MAG\_x}$,   $\tt{N\_x}$,  $\tt{AR\_x}$,  $\tt{PA\_x}$ and their respective errors are the final structural parameters obtained in Sect.  \ref{section:struct_param_estimate} from the fit on the full images.  The weighted-means $\tt{r_{e}\_x}$,  $\tt{MAG\_x}$,  $\tt{AR\_x}$ and the errors $\tt{r_{e}\_ERR\_x}$, $\tt{MAG\_ERR\_x}$, $\tt{AR\_ERR\_x}$ of the effective radii in pixels,  magnitudes and axis ratios in the $\mathrm{x}$-band are used to compute the circularised effective radii in kpc, 
\begin{equation}
\tt{RE\_x} = \tt{r_{e}\_x} \times \sqrt{\tt{AR\_x}} \times \mathrm{pixel\_scale} \times \mathrm{kpc\_per\_arcsec} \ ,
\end{equation}
the average surface brightnesses within that radius,
\begin{eqnarray}
\label{weighted_average_formula}
\nonumber \tt{MU\_x} &=&  \tt{MAG\_x} + 2.5 \log{(2 \pi)} + 5 \log{(\tt{RE\_x}}  / \mathrm{kpc\_per\_arcsec}) \\
&-& 10 \log{(1 + z_{\mathrm{cluster}})} \ ,
\end{eqnarray}
and their respective errors $\tt{RE\_ERR\_x}$ and $\tt{MU\_ERR\_x}$.


%
%

\end{document}